\documentclass{PoS}

\usepackage{graphicx}
\usepackage{comment} 
\usepackage{rotating}
\usepackage{epsfig}
\usepackage{amssymb,amsmath,amsfonts}
\usepackage{bm}
\usepackage{paralist}
\usepackage{cite}
\usepackage{booktabs}
\usepackage{multirow}
\def\OMIT#1{{}}

\newcommand{\beq}{\begin{equation}}
\newcommand{\eeq}{\end{equation}}

\newcommand{\bea}{\begin{eqnarray}}
\newcommand{\eea}{\end{eqnarray}}

\newcommand{\bee}{\begin{enumerate}}
\newcommand{\eee}{\end{enumerate}}

\newcommand{\bef}{\begin{figure}}
\newcommand{\eef}{\end{figure}}

\newcommand{\bei}{\begin{itemize}}
\newcommand{\eei}{\end{itemize}}

\newcommand{\benn}{\begin{displaymath}}
\newcommand{\eenn}{\end{displaymath}}
\newcommand{\ket}[1]{| #1 \rangle}                     
\newcommand{\bra}[1]{\langle #1 \, |}                  

\title{Nucleon axial form factors from two-flavour Lattice QCD}

\ShortTitle{Nucleon axial form factors from two-flavour Lattice QCD}

\author{\speaker{P.~M.~Junnarkar}$^a$, S.~Capitani$^a$, D.~Djukanovic$^a$, G.~von~Hippel$^b$, J.~Hua$^b$, B.~J{\"a}ger$^c$, H.~B.~Meyer$^{a,b}$, T.~D.~Rae$^a$, H.~Wittig$^{a,b}$\\
        	$^a$Helmholtz-Insititut Mainz, $^b$PRISMA Cluster of Excellence and Institut f{\"u}r Kernphysik,\\
	Johannes Gutenberg-Universität Mainz, D-55099 Mainz, Germany.\\
        $^c$Department of Physics, College of Science, Swansea University, SA2 8PP Swansea, UK. \\
         E-mail: \email{junnarka@kph.uni-mainz.de},
         \email{capitan@kph.uni-mainz.de},
         \email{djukanov@uni-mainz.de},
         \email{hippel@kph.uni-mainz.de},
         \email{hua@kph.uni-mainz.de},
         \email{jaeger@kph.uni-mainz.de},
         \email{meyerh@kph.uni-mainz.de},
         \email{thrae@uni-mainz.de},
         \email{wittig@kph.uni-mainz.de}}

\abstract{We present preliminary results on the axial form factor $G_A(Q^2)$ and the induced pseudoscalar form factor $G_P(Q^2)$ of the nucleon. A systematic analysis of the excited-state contributions to form factors is performed on the CLS ensemble `N6' with $m_\pi = 340 \ \text{MeV}$ and lattice spacing $a \sim 0.05 \ \text{fm}$. 
The relevant three-point functions were computed with source-sink separations ranging from $t_s \sim 0.6 \ \text{fm}$ to $t_s \sim \ 1.4 \ \text{fm}$. We observe that the form factors suffer from non-trivial excited-state contributions at the source-sink separations available to us. It is noted that naive plateau fits underestimate the excited-state contributions and that the method of summed operator insertions correctly accounts for these effects.
}

\FullConference{The 32nd International Symposium on Lattice Field Theory,\\
		23-28 June, 2014\\
		Columbia University New York, NY}

\begin{document}

\section{Introduction}

The nucleon matrix elements of the iso-vector axial current can be parametrised in terms of the axial form factor $G_A(Q^2)$ and the induced pseudoscalar form factor $G_P(Q^2)$, using Lorentz and CPT invariance as well as isospin symmetry, as:
\beq
 \bra{N,\bm{p}^{\prime}} A^{\mu}(x) \ket{N,\bm{p}} = e^{-i \bm{q} \cdot \bm{x}}  \bar{u}(\bm{p}^{\prime}) \big( \gamma^\mu \gamma^5 G_A(Q^2) + \gamma^5 \frac{q^\mu}{2m} G_P(Q^2)\big) u(\bm{p}),
 \eeq
where $u(\bm{p})$ is a Dirac spinor with momentum $\bm{p}$,  $\ Q^2 =  - q^2 = \bm q^2 - (E_{\bm p'}-E_{\bm p})^2$ and  $\bm{q} = \bm{p^\prime} - \bm{p}$.
The axial form factor at zero momentum transfer corresponds to the axial charge $g_A$ and is measured to a high accuracy in neutron beta decay experiments \cite{Beringer:1900zz}.  
Phenomenologically,  $G_A(Q^2)$ is represented by a  dipole as:
\beq
G_A (Q^2) = g_A \bigg / \bigg(1 + \frac{Q^2}{M^2_A}\bigg)^2 , \quad \quad \langle r^2_A\rangle = -\frac{6}{g_A} \frac{\partial G_A(Q^2)}{\partial Q^2}\bigg|_{Q^2 = 0},
\eeq
where $M_A$ is the axial pole mass, which can be related to the axial radius of the nucleon, $r_A$. 
The axial form factor $G_A(Q^2)$ is experimentally accessible via charged pion electroproduction  and elastic neutrino scattering \cite{Bernard:2001rs}. The structure of the induced pseudoscalar form factor $G_P(Q^2)$ is constrained by chiral symmetry breaking to have a pion pole as:
\beq
\label{eq:GPQ}
G_P(Q^2) = G_A(Q^2) \frac{4 M_N^2 }{Q^2 + m^2_\pi}.
\eeq
Experimentally, $G_P(Q^2)$ is measured in the muon capture process on the proton \cite{Bernard:2001rs} and is the least well known of the nucleon form factors.

The lattice calculation of the axial structure observables of the nucleon is pursued by several groups \cite{Constantinou:2014tga} (and references therein), where the axial charge $g_A$, being a benchmark quantity, is of central interest. As has been shown in \cite{Capitani:2012gj,Syritsyn:2014saa,Jager:2013kha},  the underestimation of $g_A$ is attributed to poorly understood systematic effects. Excited-state contributions to form factors, being one such effect, are found to be non-trivial~\cite{vonHippel:2014hla} and motivates the need for a careful study.  Here, we perform an analysis of the axial form factor with the focus on excited-state contributions. Our simulations use non-perturbatively $\mathcal{O}(a)$ improved Wilson fermions in $N_f = 2$ QCD, generated as part of the CLS effort with details provided in \cite{Jager:2013kha}.

\section{Lattice systematics}
The evaluation of axial matrix elements involves  constructing a ratio of three-point and two-point functions of the nucleon, defined as \cite{Alexandrou:2008rp}:
\beq 
\label{eq:ratio}
R_{\gamma_{\mu}\gamma_{5}}(\bm{q},t,t_s) \equiv \frac{C_{3,\gamma_{\mu}\gamma_{5}}(\bm{q},t,t_s)}{C_2(0,t_s)} \sqrt{ \frac{C_2(\bm{q},t_s-t)C_2(\bm{0},t)C_2(\bm{0},t_s)}{C_2(\bm{0},t_s-t)C_2(\bm{q},t)C_2(\bm{q},t_s)}}.
\eeq
The  three-point and two-point functions are computed as follows:
\beq
C_3(t) =  \sum_{\bm{x}, \bm{y}} e^{i \bm{q} \cdot \bm{y} } \Gamma_{\beta \alpha}  \langle N_{\alpha}(\bm{x}) \mathcal{O}_{\gamma_{\mu}\gamma_{5}}(\bm{y},t)\bar{N}_\beta (0) \rangle,  \quad \quad C_2(t) =  \sum_{\bm{x}}  e^{i \bm{q} \cdot \bm{x} } \Gamma_{\beta \alpha} \langle N_{\alpha}(\bm{x}) \bar{N}_\beta (0)  \rangle,
\eeq
where $N_\alpha(x)$ is a nucleon interpolating operator and $\Gamma_{\alpha \beta}$ is a projection matrix  which projects $N_\alpha$ to have the correct parity. The polarisation of the nucleon is chosen to be in the $z$-direction as $\Gamma = \frac{1}{2} (1+ \gamma_0) (1 + i \gamma_5 \gamma_3)$. The three-point function represents a nucleon interacting with an axial current. The relevant diagram is computed by contracting the fixed-sink sequential propagator with the operator and ordinary propagator. The kinematics is such that the transferred momentum $\bm{q}$ is injected  at the operator while the nucleon at the sink is at rest.  To improve the overlap of the nucleon interpolating operator with the ground state, Gaussian smearing \cite{Gusken:1989ad} is applied to quark fields supplemented with APE smeared links \cite{Albanese:1987ds} at both source and sink.

\section{Form factor extraction}
The asymptotic behaviour of the three-point and two-point functions yields the ground state form factors from eq.~(\ref{eq:ratio}) as:
\beq
R^0_{\gamma_{0}\gamma_{5}}(\bm{q},t,t_s) \xrightarrow[t_s \rightarrow \infty]{} \frac{q_3}{\sqrt{2 E_q(m+E_q)}} \bigg( G^{\rm{bare}}_A(Q^2) + \frac{m-E_q}{2m} G^{\rm{bare}}_P(Q^2) \bigg),
\eeq
\beq
\label{eq:componentratio}
R^0_{\gamma_{k}\gamma_{5}}(\bm{q},t,t_s)  \xrightarrow[t_s \rightarrow \infty]{}  \frac{i}{\sqrt{2 E_q(m+E_q)}} \bigg( (m+E_q)  G^{\rm{bare}}_A(Q^2) \delta_{3k} - \frac{G^{\rm{bare}}_P(Q^2)}{2m} q_3 q_k \bigg).
\eeq
It is clear from eq.~(\ref{eq:componentratio}) that the ratio $R_{\gamma_3 \gamma_5}$ provides direct access to $G^{\rm{bare}}_A(Q^2)$ at appropriate kinematics i.e at $q_3 = 0$. $G^{\rm{bare}}_P(Q^2)$ can also be extracted thereafter from $R_{\gamma_3 \gamma_5}$ by including the $q_3 \neq 0 $ channels. The bare form factors are renormalised as, $G_X(Q^2) = Z_A \ G^{\rm{bare}}_X(Q^2)$~\footnote{A mass dependent improvement term is negligible in the mass range considered here.}, where $Z_A$ was determined non-perturbatively in \cite{DellaMorte:2008xb}.
The ratio as defined in eq.~(\ref{eq:ratio}) contains excited-state contributions coming from two-point and three-point functions given as:
\beq
R(\bm{q},t,t_s) = R^{0}(\bm{q},t,t_s) \bigg(1 + \mathcal{O}(e^{-\Delta t}) + \mathcal{O}(e^{-\Delta^{\prime} (t_s-t)})\bigg),
\eeq
where $\Delta$ and $\Delta^{\prime}$ are the excited-state energy gaps of the initial and final state nucleon.
To study the effects of these contributions systematically, we focus on ensemble `N6', where we have a particularly large data sample at our disposal.  On this ensemble,  the three-point function is computed for source-sink separations ranging from $t_s \sim 0.6 \ \text{fm} $ to $ t_s \sim  1.4 \ \text{fm}$ with a sample size of $N_{\rm{cfg}}=946$ gauge configurations. We also employ three different methods to investigate these contributions, i.e.

\bee
\item Fit a plateau for $t_s \sim 1.1 \ \text{fm}$ source-sink separation to a constant.\footnote{For fitting a plateau, the source-sink separation of $t_s \sim 1.1 \ \text{fm}$ is the largest available which yields meaningful uncertainties on the fit. Data on $t_s > 1.1 \ \text{fm}$ is not considered as the uncertainties are higher.}
\item Construct summed ratios \cite{Maiani:1987by,Brandt:2011sj} as:
\beq
S(t_s) = \sum^{t_s}_{t=0} R(\bm{q},t,t_s) \rightarrow c(\Delta,\Delta^{\prime}) + t_s \bigg(G^{\rm{bare}}_{A,P} + \mathcal{O}(e^{-\Delta t_s}) + \mathcal{O}(e^{-\Delta^{\prime} t_s}) \bigg)
\eeq
and extract the form factors as the slope of $S(t_s)$ for various source-sink separations.
\item Perform simultaneous two-state fits, where one makes an explicit ansatz about the nature of the energy gaps of excited-states and the data is fitted to the following:
\beq R_{\gamma_3 \gamma_5}(Q^2,t,t_s) = G^{\rm{bare}}_X(Q^2) + c_1 e^{-\Delta t} + c_2 e^{-\Delta^{\prime} (t_s - t )}, \eeq
\beq \label{eq:gA-2state} R_{\gamma_3 \gamma_5}(Q^2=0,t_s) = g^{\rm{bare}}_A + c_1 e^{-\Delta t_s/2}.\eeq
\eee
The two-state fit for the axial charge simplifies at $Q^2 = 0$, as one has $\Delta^\prime = \Delta$  and the $t$ dependence of $g^{\rm{bare}}_A$ is eliminated by averaging over five time slices  around $t = t_s/2 $. \footnote{For source-sink separations that are odd in lattice units, plateaus are fitted to data at $\big((t_s - 1)/2 \pm 2\big)$.} 

\section{Analysis of the axial form factor $G_A(Q^2)$}
\bef[t!]
\hspace{-1.0cm}
\includegraphics[width=0.55\linewidth]{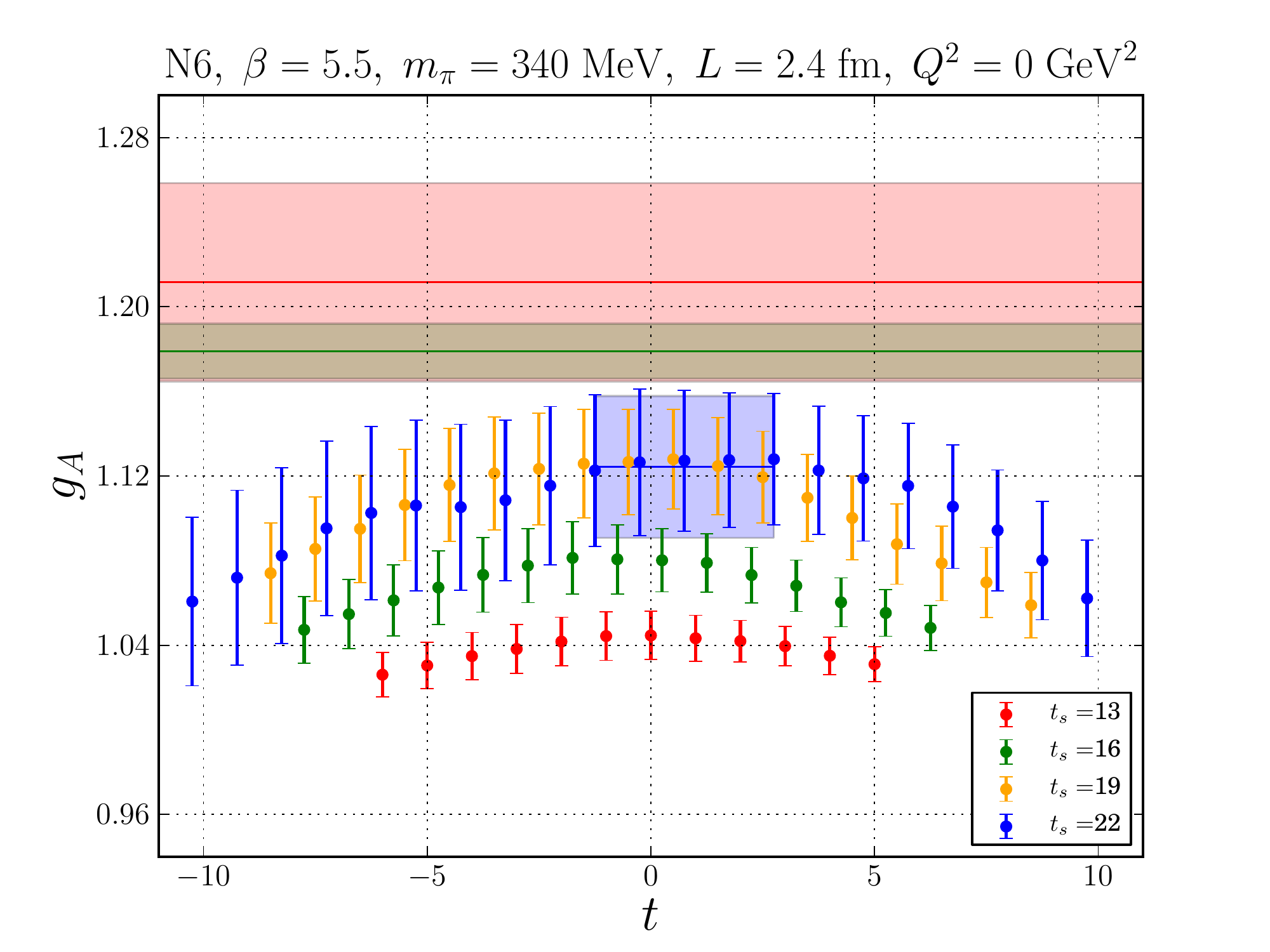}
\includegraphics[width=0.55\linewidth]{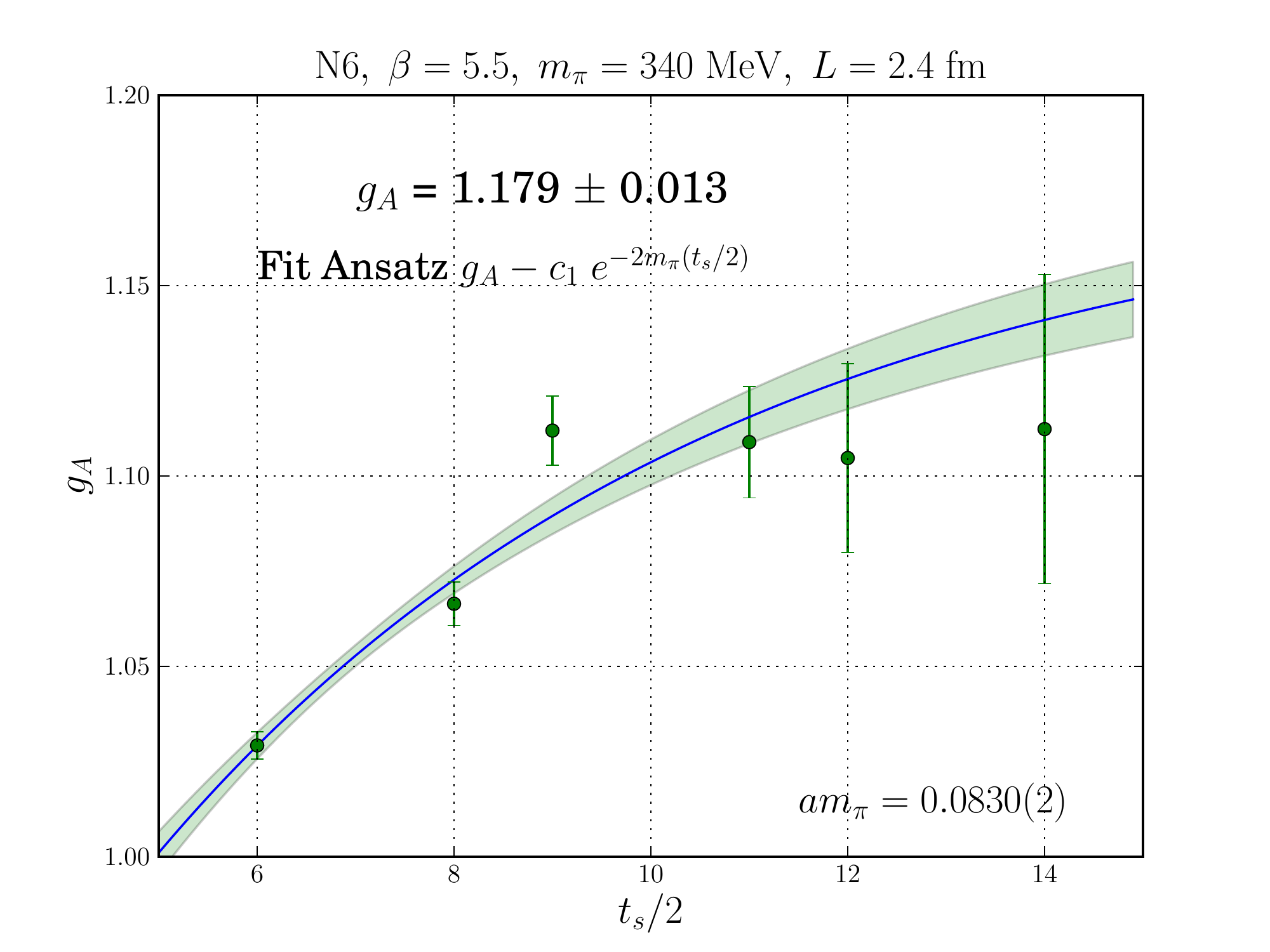}
\caption{\label{fig:gA1} Right: Comparison of results of two-state fits (overlapping green band) with summation method (red band) and plateau fit (blue band). Left: A two-state fit to $g_A$.}
\vspace{-0.25cm}
\eef
The results for the axial charge, as presented in \cite{Jager:2013kha}, are summarised in left panel of Fig~\ref{fig:gA1}.
It is clear that the results of the plateau and summation method fits do not mutually agree. In order to obtain a deeper insight, we also include data with $t_s \ge 1.1 \ \text{fm}$ and fit plateaus to each of the source-sink separations. The results are plotted as a function of $t_s/2$ as shown in the right panel of  Fig~\ref{fig:gA1}. We then fit this data with a two-state fit as in eq.~(\ref{eq:gA-2state}) with the gap fixed to $\Delta = 2 m_\pi$. The comparison of the results is presented in the left panel of Fig~\ref{fig:gA1}, and the two-state fit  (overlapping green band) agrees well with the summation method indicating that the summation method correctly takes into account the excited-state contributions. Furthermore, it does so without making any assumption about the nature of gaps.
\bef[t!]
\centering
\includegraphics[width=0.55\linewidth]{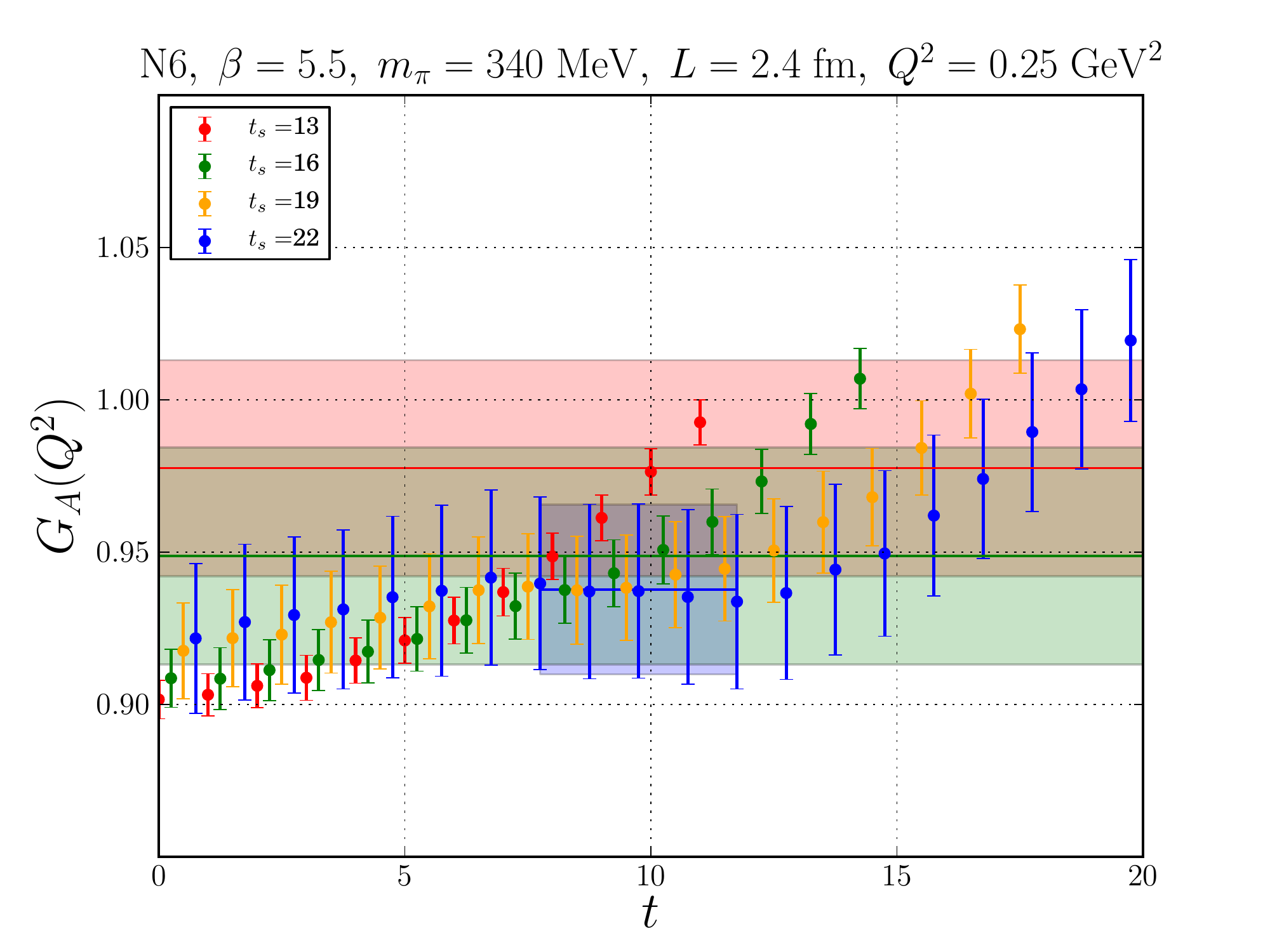}
\caption{\label{fig:GAqsq1} Results at lowest non-zero $Q^2$. Plateau fit with blue band, summation method with red band  and two-state fit with green band.}
\vspace{-0.2cm}
\eef
\bef[t!]
\hspace{-1.0cm}
\includegraphics[width=0.55\linewidth]{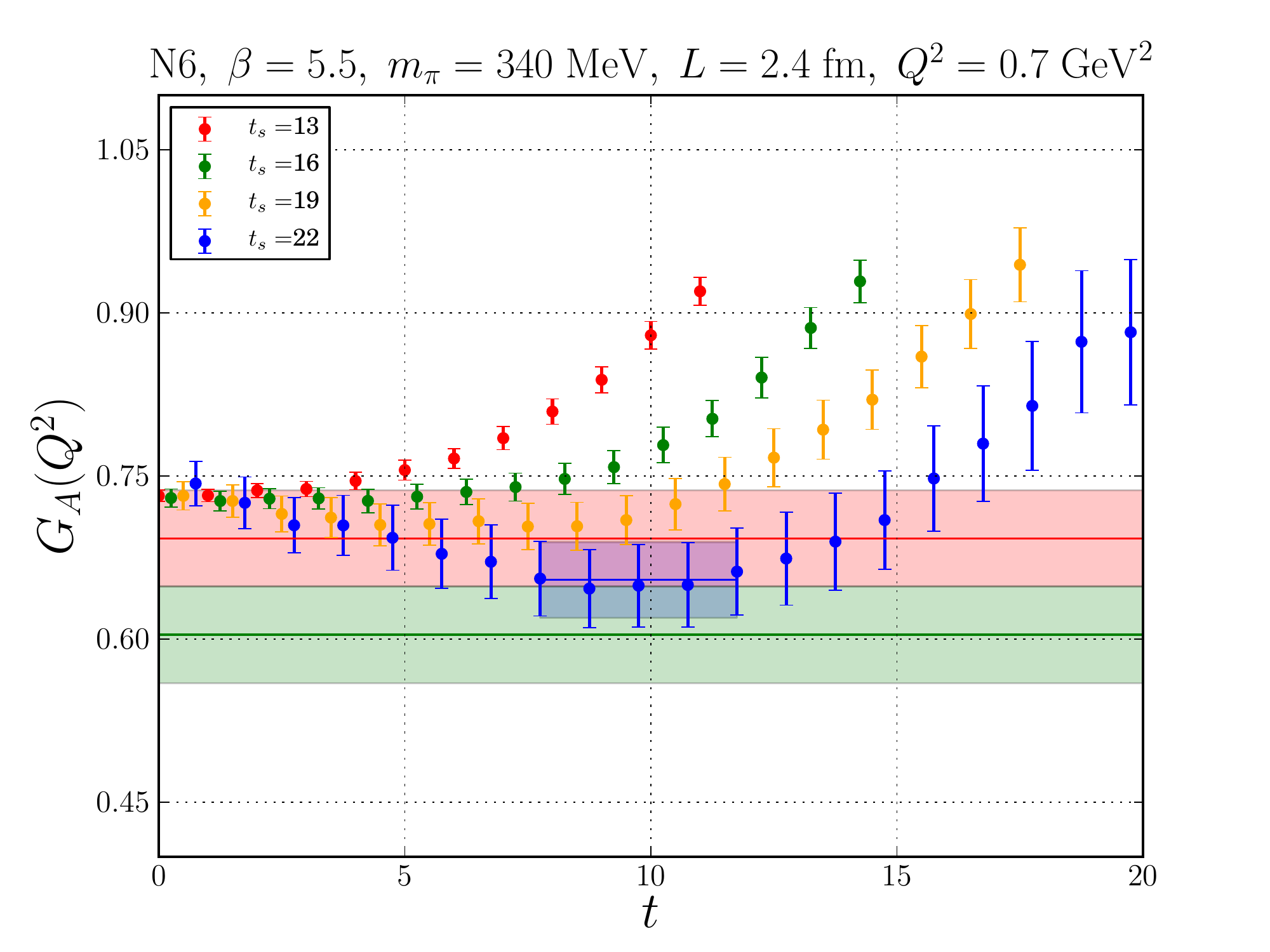}
\includegraphics[width=0.55\linewidth]{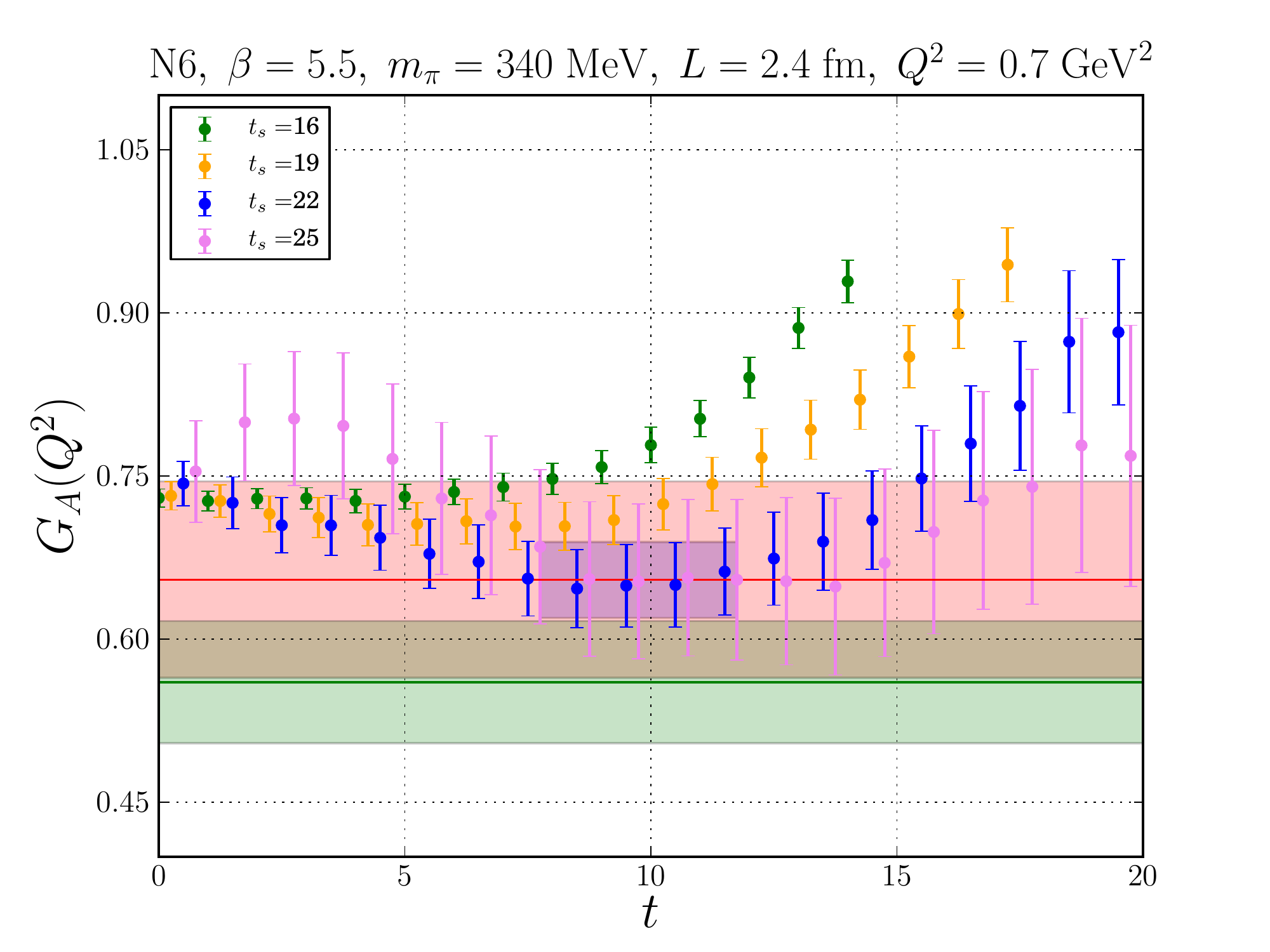}
\caption{\label{fig:GAqsq3} Results at $Q^2 = 0.7 \ \text{GeV}^2$. Left: Results with inclusion of lowest $t_s$.  Right: Results with elimination of lowest $t_s$. Colour scheme for bands is same as in Fig 2.}
\vspace{-0.2cm}
\eef

The results at the lowest non-zero $Q^2$ are presented in Fig~\ref{fig:GAqsq1}.  As can be seen, even though the results for plateau, summation and two-state fits overlap, the absence of a plateau-like behaviour for $t_s \sim 0.6-0.8 \ \text{fm}$ (red and green data points) is a clear indication that the ground-state has not been reached. Further, for source-sink separations of $1.1 \ \text{fm}$ (blue data points), the uncertainties are high enough, that it is difficult to determine the extent of the excited-state contributions. It is also noted that at the lowest non-zero $Q^2$, the available degenerate momentum channels are lowest and therefore the gain in statistics is less compared to higher momentum channels. Hence, we believe that the excited-state contributions can be more clearly exposed with improved statistics.

 At higher $Q^2$, the excited-state contributions are obvious from the data, as is shown in Fig~\ref{fig:GAqsq3}. The summation method fit, left Fig~\ref{fig:GAqsq3}, seems to describe the data well, however we note that data at the lowest $t_s$, being more accurate, dominates the fit. This is observed in the right panel of Fig~\ref{fig:GAqsq3}, where after eliminating the lowest $t_s$, the summation method results exhibit a downward trend (albeit with higher uncertainties) more inline with the two-state fits. The results of various fits at all $Q^2$ is summarised in the left panel of Fig~\ref{fig:GAGP}.
\section{Induced pseudoscalar form factor $G_P(Q^2)$}
\bef[t!]
\centering
\includegraphics[width=0.55\linewidth]{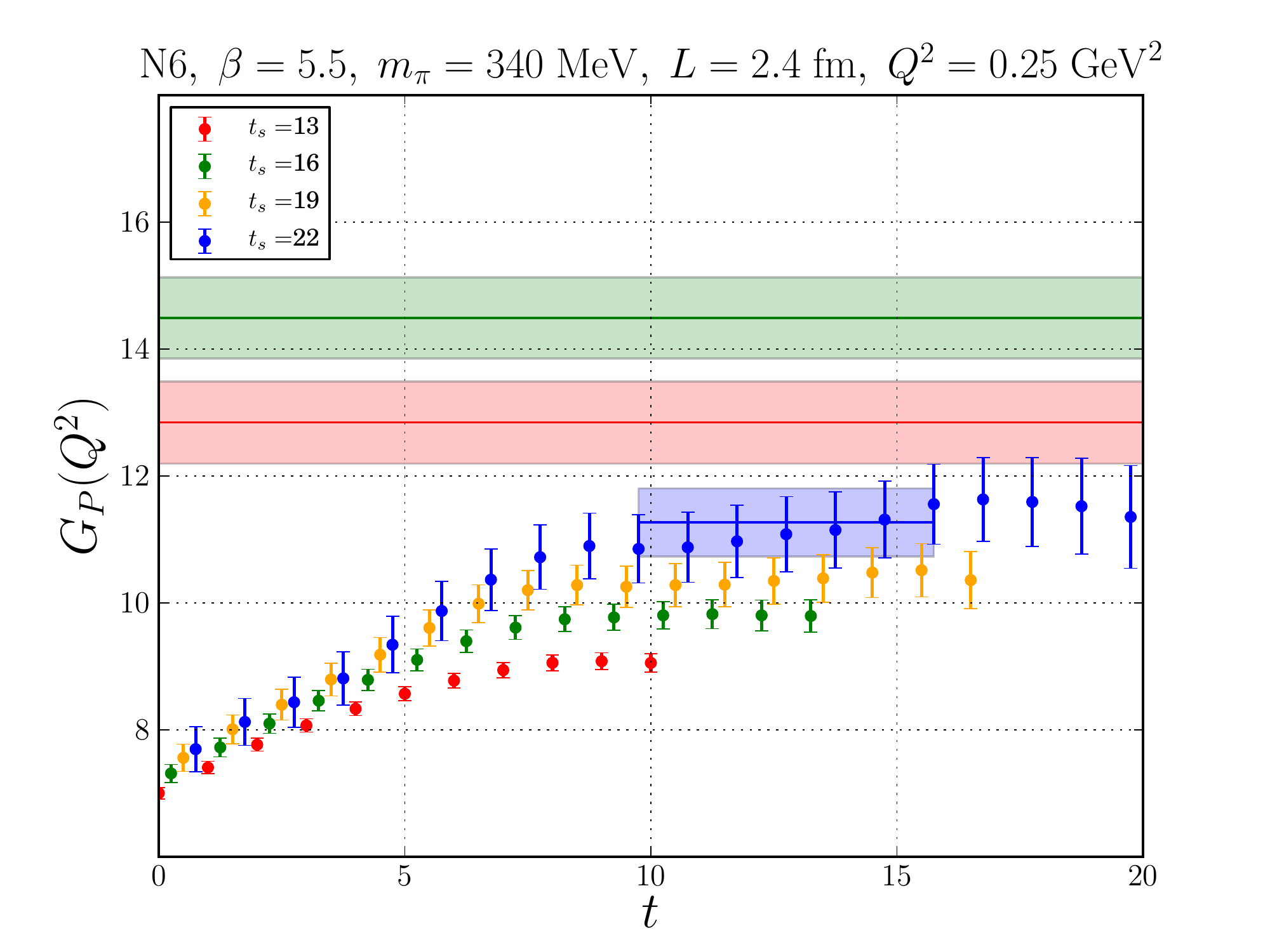}
\caption{\label{fig:GP} Results for $G_P(Q^2)$ at lowest non-zero $Q^2$. Colour scheme for bands is same as in Fig 2.}
\vspace{-0.2cm}
\eef
\bef[t!]
\hspace{-1.0cm}
\includegraphics[width=0.55\linewidth]{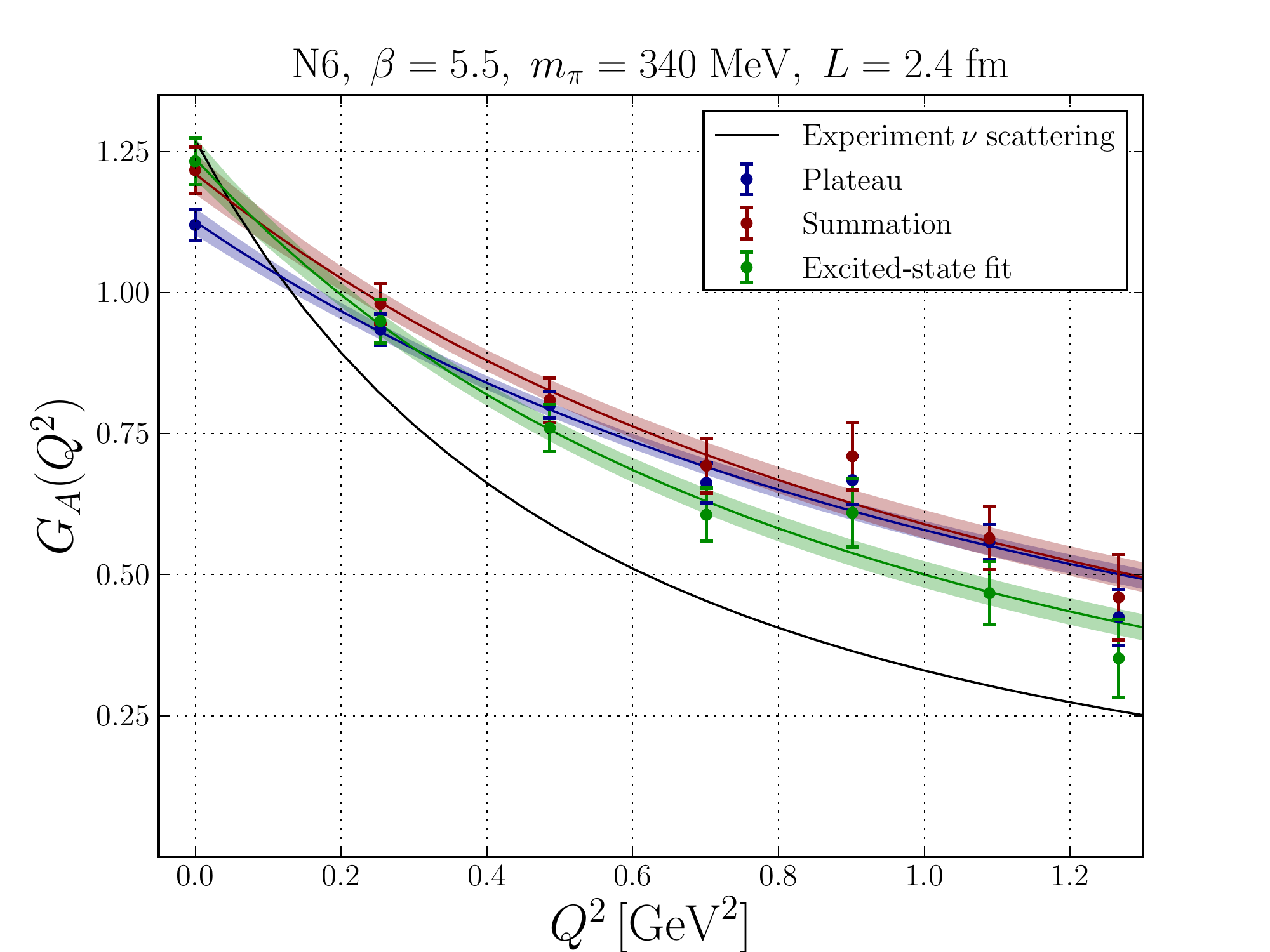}
\includegraphics[width=0.55\linewidth]{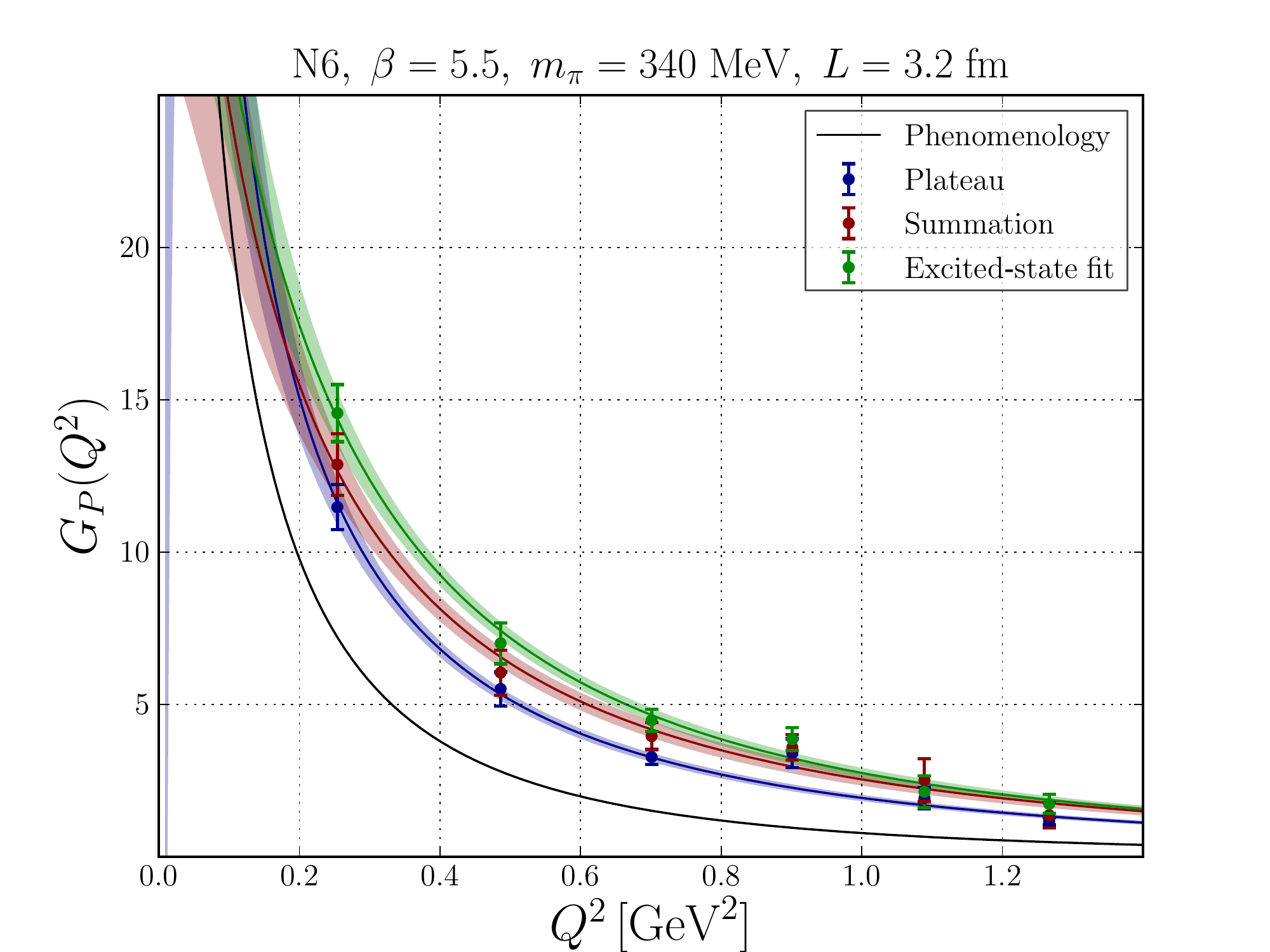}
\caption{\label{fig:GAGP} Left:  Results of various fits for $G_A(Q^2)$  Right: Results of various fits for $G_P(Q^2)$}
\vspace{-0.2cm}
\eef

The analysis of $G_P(Q^2)$ is more complicated than $G_A(Q^2)$, because with our kinematic setup we cannot directly access $G_P(Q^2)$. The data for $G_P(Q^2)$ is obtained from $R_{\gamma_3 \gamma_5}$ for the kinematics with $q_3 \neq 0$  which presumes the knowledge of $G_A(Q^2)$.  In the results presented, the data for $G_P(Q^2)$ were obtained by subtracting the data for $G_A(Q^2)$ from $R_{\gamma_3 \gamma_5}$.The results for $G_P(Q^2)$ are presented in Fig~\ref{fig:GP} and in the right panel of Fig~\ref{fig:GAGP}.  At the lowest non-zero $Q^2$, the results for the various fits methods do not agree, indicating non-trivial excited-state contributions. At this point, we note that the excited-state contributions coming from $G_A(Q^2)$ are not accounted for, which motivates us to perform a more thorough analysis for $G_P(Q^2)$.  We also verify the presence of the pion pole, by fitting the data for each method to the ansatz in eq.~(\ref{eq:GPQ}).
We emphasise that the results presented are preliminary and a more detailed analysis will be presented in a later publication. 

\section{Conclusions and outlook}
We have presented preliminary results for the axial form factor $G_A(Q^2)$ and induced pseudoscalar form factor $G_P(Q^2)$. We find that for the case of axial charge, naive plateau fits underestimate the excited-state contributions and that the summation method correctly accounts for these effects without making any assumption about the nature of gaps.  At non-zero momenta, we observe that the data at lowest $t_s$ is responsible for underestimating the excited-state contributions. Therefore, to obtain a more correct estimate, we typically discard the data at the lowest $t_s$. We note that in all cases the two-state fits are used only as a tool to get an insight in the behaviour of excited-states, due to the fact that assumptions are made about the nature of the gaps.

There are several avenues that are currently being explored, which will be discussed in a later publication.  In extracting $G_P(Q^2)$,  the excited-state contributions coming from $G_A(Q^2)$  can be avoided by estimating the combined matrix elements from $R_{\gamma_3 \gamma_5}$ directly from one of the three fitting methods. Our results presented here were  obtained using the unimproved axial current and can be $\mathcal{O}(a)$ improved, as the relevant coeffcients are known non-perturbatively. The pseudoscalar form factor is related to axial form factors through the PCAC relation and allows us to study the Goldberger-Treiman relation. The chiral and continuum extrapolation of the axial radius $ r_A$ obtained from $G_A(Q^2)$ also remains to be explored and will be presented in detail in a later publication. \\

{\noindent {\bf Acknowledgments:}
\noindent
Our calculations were performed on the ``Wilson'' HPC Cluster at the Institute for Nuclear Physics, University of Mainz. We thank 
Christian Seiwerth for technical support. We are grateful for computer time allocated to project HMZ21 on the BG/Q ``JUQUEEN'' computer at NIC, J\"ulich. This work was granted access to the HPC resources of the Gauss Center for Supercomputing at Forschungzentrum J\"ulich, Germany, made available within the Distributed European Computing Initiative by the PRACE-2IP, receiving funding from the European Community's Seventh Framework Programme (FP7/2007-2013) under grant agreement RI-283493. This work was supported by the DFG via SFB 1044 and grant HA 4470/3-1. We are grateful to our colleagues within the CLS initiative for sharing ensembles.

\bibliographystyle{JHEP-2-notitle}
\bibliography{axial.v1}
\end{document}